# Distributed Cooperative Spectrum Sensing in Mobile Ad Hoc Networks with Cognitive Radios

F. Richard Yu, Helen Tang, Minyi Huang, Peter Mason, and Zhiqiang Li

In cognitive radio mobile ad hoc networks (CR-MANETs), secondary users can cooperatively sense the spectrum to detect the presence of primary users. In this chapter, we propose a fully distributed and scalable cooperative spectrum sensing scheme based on recent advances in consensus algorithms. In the proposed scheme, the secondary users can maintain coordination based on only local information exchange without a centralized common receiver. We use the consensus of secondary users to make the final decision. The proposed scheme is essentially based on recent advances in consensus algorithms that have taken inspiration from complex natural phenomena including flocking of birds, schooling of fish, swarming of ants and honeybees. Unlike the existing cooperative spectrum sensing schemes, there is no need for a centralized receiver in the proposed schemes, which make them suitable in distributed CR-MANETs. Simulation results show that the proposed consensus schemes can have significant lower missing detection probabilities and false alarm probabilities in CR-MANETs. It is also demonstrated that the proposed scheme not only has proven sensitivity in detecting the primary user's presence, but also has robustness in choosing a desirable decision threshold.

F. Richard Yu
Department of Systems and Computer Engineering, Carleton University, Ottawa, ON, Canada e-mail: richard_yu@carleton.ca

Helen Tang
Defense R&D Canada - Ottawa, ON, Canada e-mail: helen.tang@drdc-rddc.gc.ca

Minyi Huang
School of Mathematics and Statistics, Carleton University, Ottawa, ON, Canada e-mail: mhuang@math.carleton.ca

Peter Mason
Defense R&D Canada - Ottawa, ON, Canada e-mail: peter.mason@drdc-rddc.gc.ca

Zhiqiang Li
Department of Systems and Computer Engineering, Carleton University, Ottawa, ON, Canada e-mail: zlia@sce.carleton.ca

# 1 Introduction

Recently, there has been tremendous interest in the field of cognitive radio (CR), which has been introduced in [1]. CR is an enabling technology that allows unlicensed (secondary) users to operate in the licensed spectrum bands. This can help to overcome the lack of available spectrum in wireless communications, and achieve significant improvements over services offered by current wireless networks. It is designed to sense the changes in its surroundings, thus learns from its environment and performs functions that best serve its users. This is a very crucial feature of CR networks which allow users to operate in licensed bands without a license [2]. To achieve this goal, spectrum sensing is an indispensable part in cognitive radio.

There are three fundamental requirements for spectrum sensing. In the first place, the unlicensed (secondary) users can use the licensed spectrum as long as the licensed (primary) user is absent at some particular time slot and some specific geographic location. However, when the primary user comes back into operation, the secondary users should vacate the spectrum instantly to avoid interference with the primary user. Hence, a first requirement of cognitive radio is that the continuous spectrum sensing is needed to monitor the existence of the primary user. Also, since cognitive radios are considered as lower priority and they are secondary users of the spectrum allocated to a primary user, the second fundamental requirement is to avoid the interference to potential primary users in their vicinity [3, 38]. Furthermore, primary user networks have no requirement to change their infrastructure for spectrum sharing with cognitive radios. Therefore, the third requirement is for secondary users to be able to independently detect the presence of primary users.

Taking those three requirements into consideration, such spectrum sensing can be conducted non-cooperatively (individually), in which each secondary user conducts radio detection and makes decision by itself. However, the sensing performance for one cognitive user will be degraded when the sensing channel experiences fading and shadowing [4, 26]. In order to improve spectrum sensing, several authors have recently proposed collaboration among secondary users [3, 5–7], which means a group of secondary users perform spectrum sensing by collaboration. As the result, it shows that collaboration may enhance secondary spectrum access significantly [5].

Our research is focused on the distributed cooperative spectrum sensing (DCSS) in cognitive radio, and more precisely, the distributed cooperative schemes of spectrum sensing in a Cognitive Radio Mobile Ad-hoc NETworks (CR-MANETs).

In the first place, at present, distributed cooperative detection problems are discussed in [6, 8–10, 23]. In a typical wireless distributed detection problem, each sensor or secondary user individually forms its own discrete messages based on its local measurement and then reports to a fusion center via wireless reporting channels. In certain models [10], however, there is in general no direct communication among the sensors.

A sensor may indirectly obtain information about other sensors, but this is achieved by feedback from a common fusion center. Nevertheless, a centralized fusion center may not be available in some CR-MANETs. Moreover, as indicated in [11], gathering the entire received data at one place may be very difficult under practical communication constraints. In addition, authors of [4] study the reporting channels between the cognitive users and the common receiver. The results show that there are limitations for the performance of cooperation when the reporting channels to the common receiver are under deep fading.

Based on recent advances in consensus algorithms [12], we propose a new scheme in distributed cooperative spectrum sensing called distributed consensus-based cooperative spectrum sensing (DCCSS).

The main contributions of this work include:

- We propose a consensus-based spectrum sensing scheme, which is a fully distributed and scalable scheme. Unlike many existing schemes [29, 32, 60], there is no need for a common receiver to do data fusion and to reach the final decision. Since it is rare to have a centralized node in MANETs, in the proposed scheme, a secondary user needs only to setup local interactions without centralized information exchange [17, 18].
- Unlike most decision rules, such as OR-rule or n-out-of-N, adopted in existing spectrum sensing schemes, we use consensus from secondary users. The proposed scheme has self-configuration and self-maintenance capabilities,
- Since the CR paradigm imposes human-like characteristics (e.g., learning, adaptation and cooperation) in wireless networks, the bio-inspired consensus algorithm used in this work can provide some insight into the design of future CR-MANETs.

Extensive simulation results illustrate the effectiveness of the proposed scheme. It is shown that the proposed scheme can have both lower missing detection probability and lower false alarm probability compared to the existing schemes. In addition, it is able to make better detection when secondary users undergo worse fading (lower average SNR). Last but not the least, with the help of this scheme, a fixed threshold is feasible, which can take active effect in different fading channels.

The rest of the chapter is organized as follows. Section 2 describes the research background of this research, which includes spectrum sensing in cognitive radios, cooperative spectrum sensing, and centralized/distributed cooperative spectrum sensing. Section 3 presents system models, spectrum sensing model, fixed/random graphs theories and consensus notions. In Section 4, the distributed consensus-based cooperative spectrum sensing scheme is proposed based on fixed graphs, together with the network models. Going further, the distributed consensus-based cooperative spectrum sensing scheme based on random graphs is described in Section 5. In Section 6, the simulation results and discussions are presented. Finally, we conclude this chapter in Section 7.

## 2 Background

This section is intended to cover the topics regarding the research background. They include the introduction of cognitive radio, functionalities of cognitive radio, differences of individual spectrum sensing and cooperative spectrum sensing, followed by the introduction of centralized distributed cooperative spectrum sensing and distributed consensus-based cooperative spectrum-sensing.

### 2.1 Introduction of Spectrum Sensing in Cognitive Radio

The idea of cognitive radio is first presented officially in an article by Joseph Mitola and Gerald Q. Maguire, Jr. [13]. It is a novel approach in wireless communications that Mitola later describe in his PhD dissertation as:

"The point in which wireless Personal Digital Assistants (PDAs) and the related networks are sufficiently computationally intelligent about radio resources and related computer-to-computer communications to detect user communications needs as a function of use context, and to provide radio resources and wireless services most appropriate to those needs."

It is thought of as an ideal goal towards which a software-defined radio platform should evolve: a fully reconfigurable wireless black-box that automatically changes its communication variables in response to network and user demands.

The above citation originates from the following fact. On one hand, the growing number of wireless standards is occupying more and more naturally limited frequency bandwidth for exclusive use as licensed bands. However, large part of licensed bands are unused for what concerns a large amount of both time and space: even if a particular range of frequencies is reserved for a standard, at a particular time and at a particular location it could be found free. The Federal Communication Commission (FCC) estimates that the variation of use of licensed spectrum ranges from 15% to 85%, whereas according to Defence Advance Research Projects Agency (DARPA) only the 2% of the spectrum is in use in US at any given moment. It is then clear that the solution to these problems can be found dynamically looking at spectrum as a function of time and space.

With the high demand of bit transmission rate for 4G or IMT-advanced high-speed wireless applications, there are several approaches to increase the system capacity as stated in the following equation:

$$C = n \cdot B \cdot log_2(1 + SNR) \qquad (1)$$

The first approach is using MIMO to increase $n$, so that capacity may have a gain proportionally. The second approach is trying to increase $SNR$. The third one is focusing on the bandwidth. Cognitive radio is among the third category, and thrives to fully utilize the frequency.

### 2.1.1 Functionalities of Cognitive Radios

The main functionalities of cognitive radios are [14]:

- **Spectrum Sensing (SS)**: detecting the unused spectrum and sharing it without harmful interference with other users, it is an important requirement of the cognitive Radio network to sense spectrum holes, detecting primary users is the most efficient way to detect spectrum holes. Spectrum sensing techniques can be classified into three categories:
    - Transmitter detection: cognitive radios must have the capability to determine if a signal from a primary transmitter is locally present in a certain spectrum, there are several approaches proposed:
        · Matched filter detection
        · Energy detection
        · Cyclostationary feature detection
    - Cooperative detection: refers to spectrum sensing methods where information from multiple cognitive radio users are incorporated for primary user detection.
    - Interference based detection.
- **Spectrum Management (SMa)**: Capturing the best available spectrum to meet user communication requirements. Cognitive radios should decide on the best spectrum band to meet the quality of service requirements over all available spectrum bands, therefore spectrum management functions are required for cognitive radios, these management functions can be classified as: spectrum analysis and spectrum decision.
- **Spectrum Mobility (SMo)**: is defined as the process when a cognitive radio user exchanges its frequency of operation. Cognitive radio networks target to use the spectrum in a dynamic manner by allowing the radio terminals to operate in the best available frequency band, maintaining seamless communication requirements during the transition to better spectrum.
- **Spectrum Sharing (SSh)**: providing the fair spectrum scheduling method, which is one of the major challenges in open spectrum usage is the spectrum sharing. It can be regarded to be similar to generic media access control MAC problems in existing systems.

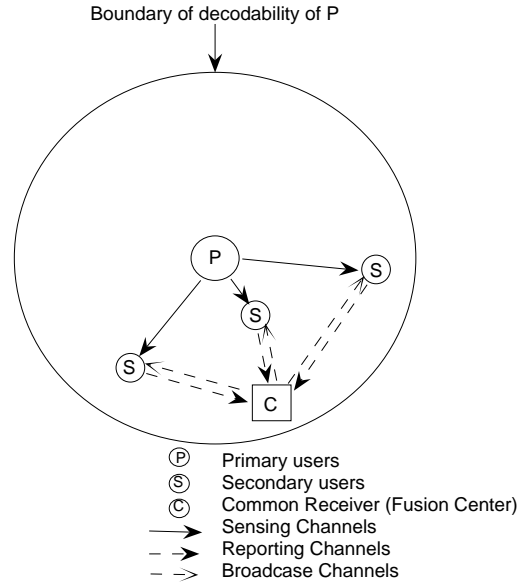

**Fig. 1:** A typical cognitive radio network.

### 2.1.2 Individual and Cooperative Spectrum Sensing

Spectrum sensing can be conducted either non-cooperatively (individually), in which each secondary user conducts radio detection and makes decision by itself, or cooperatively, in which a group of secondary users perform spectrum sensing by collaboration. No matter in which way, the common topology of such a cognitive radio network can be depicted as in Fig. 1. Individual spectrum sensing is conducted by secondary users on its own, and each user has a local observation and a local decision accordingly. Thus, in Fig. 1, each secondary user performs the spectrum sensing locally and no communication is between one another, nor is the common receiver (fusion center). In such a condition, cognitive radio sensitivity can only be improved [6] by enhancing radio RF front-end sensitivity, exploiting digital signal processing gain for specific primary user signal, and network cooperation where users share their spectrum sensing measurements. However, if the sensing channels are facing deep fading or shadowing, then affected individuals will not be able to detect the presence of the primary user, which leads to missing detection failure.

In order to improve the performance of spectrum sensing, several authors have recently proposed cooperation among secondary users [2, 4, 5, 15]. Cooperative spectrum sensing has been proposed to exploit multi-user diversity in sensing process. It is usually performed in three successive stages: sensing, reporting and broadcasting. In the sensing stage, every cognitive user performs spectrum sensing individually. This can be shown as in Fig. 1, where secondary users try to collect the signal of interest through sensing channels. In the reporting stage, all the local sensing observations are reported to a common receiver via reporting channels (see Fig. 1) and the latter will make a final decision on the absence or the presence of the primary user. Finally, the final decision is broadcasted via broadcast channels to all the secondary users concerned, which include not only the ones involved into the sensing stage, but also those that do not have sensing capabilities but want to participate into the spectrum sharing stage.

There are several advantages offered by cooperative spectrum sensing over the non-cooperative ones [5, 11, 16, 19, 24, 27–29, 32]. If a secondary user is in the condition of deep shadowing and fading, it is very difficult for a secondary user to distinguish a white space from a deep shadowing effect. Therefore, a non-cooperative spectrum sensing algorithm may not work well in this case, and a cooperative scheme can solve the problem by sharing the spectrum sensing information among secondary users. Moreover, because of the hidden terminal problem, it is very challenging for single cognitive radio sensitivity to outperform the primary user receiver by a large margin in order to detect the presence of primary users. For this reason, if secondary users spread out in the spatial distance, and any one of them detects the presence of primary users, then the whole group can gain benefit by collaboration.

Authors of [5] quantify the performance of spectrum sensing in fading environments and study the effect of cooperation. The simulation results in [5] indicate that significant performance enhancements can be achieved through cooperation. Authors of [16] study the possibility to forward the signal with higher SNR to the one on the boundary of decidability region of the primary user. The performance is evaluated under correlated shadowing and user compromise in [11]. When the exchange of observations from all secondary users to the common receiver is not applicable, authors of [19] show that it is still worth doing by cooperating a certain number of users with relatively higher SNR. Moreover, in [24], a linear-quadratic (LQ) fusion strategy is designed with the consideration of the correlation between the nodes. In order to further reduce the computational complexity, authors of [27] propose a heuristic approach so as to develop an optimal linear framework during cooperation. Sensing-throughput tradeoff is analyzed in [28] for both multiple mini-slots and multiple secondary users cooperative sensing.

**2.1.3 Centralized Cooperative Spectrum Sensing**

Although some research activities have been conducted in cooperative spectrum sensing, most of them use a common receiver (fusion center) to do data fusion for the final decision whether or not the primary user is present. However, a common receiver may not be available in some CR-MANETs. Moreover, as indicated in [11], gathering the entire received data at one place may be very difficult under practical communication constraints. In addition, authors of [4] study the reporting channels between the cognitive users and the common receiver. The results show that there are limitations for the performance of cooperation when the reporting channels to the common receiver are under deep fading. In summary, the use of a centralized fusion center in CR-MANETs may have the following problems (see Fig. 1):

- Every secondary user needs to join/establish the connection with the common receiver, which requires a network protocol to implement.
- Some secondary users need a kind of relay routes to reach the common receiver if they are far away from the latter.
- Communication errors or packet drops can affect the performance of such a network if more users have worse reporting channels (e.g. Rayleigh Fading) to reach the common receiver.
- There should be a reliable wireless broadcast channel [20, 22, 61] for the common receiver to inform each of every user once there is a decision made.
- The current centralized network does not fit for the average calculation of all the estimated sensing energy levels, because it requires the common receiver to correctly receive all the local estimated sensing results. Otherwise, the decision precision can not be guaranteed.

## 2.2 Mobile Ad Hoc Networks

In recent years, MANETs have become a popular subject because of their self-configuration and self-organization capabilities. Each device in a MANET is free to move independently in any direction, and will therefore change its links to other devices frequently. Wireless nodes can establish a dynamic network without the need of a fixed infrastructure. A node can function both as a network router for routing packets from the other nodes and as a network host for transmitting and receiving data. MANETs are particular useful when a reliable fixed or mobile infrastructure is not available. Instant conferences between notebook PC users, military applications, emergency operations, and other secure-sensitive operations are important applications of MANETs due to their quick and easy deployment.

**2.2.1 Self-organization of MANETs**

Due to the lack of centralized control, MANETs nodes cooperate with each other to achieve a common goal [30,33]. The major activities involved in self-organization are neighbor discovery, topology organization, and topology reorganization. Through periodically transmitting beacon packets, or promiscuous snooping on the channels, the activities of neighbors can be acquired. Each node in MANETs maintains the topology of the network by gathering the local or entire network information. MANETs need to update the topology information whenever the networks change such as participation of new node, failure of node and links, etc. Therefore, self-organization is a continuous process that has to adapt to a variety of changes or failures.

## 2.3 Distributed Consensus-based Cooperative Spectrum Sensing Scheme

In this work, we will present a distributed consensus-based cooperative spectrum sensing scheme without using a common receiver. Our scheme is based on recent advances in consensus algorithms [12], or more precisely, bio-inspired mechanisms, which have become important approaches to handle complex communication networks [34–36, 39]. An important motivational background of this area is initially related to the study of complex natural phenomena including flocking of birds, schooling of fish, swarming of ants and honeybees, among others (see the survey [37]). The investigation of such biological systems has generated fundamental insights into understanding the relation between group decision making at the higher level and the individual animals' communication at the lower level [31, 40–44, 62], and in fact consensus seeking in animal colonies is vital for group survival [44]. Such collective animal behavior has motivated many effective yet simple control algorithms for the coordination of multi-agent systems in engineering. Recently, consensus problems have played a crucial role in spacial distributed control models [12, 21, 45], wireless sensor networks [46], and stochastic seeking with noise measurement [47]. Since these algorithms are usually constructed based on local communication of neighboring agents, they have low implementation complexity and good robustness, and the overall system may still function when local failure occurs.

The main highlights of this scheme are as follows.

- It is a fully distributed and scalable scheme. Unlike the existing schemes [29, 32, 60], there is no need for a common receiver to do the data fusion for the final decision. A secondary user only needs to set up neighborhood with those users having desired channel characteristics, such as Line of Sight ones, or even with probabilistic link failures.

- Unlike most decision rules, such as OR-rule or 1-out-of-N, adopted in the existing schemes, we use the consensus of secondary users to make the final decision. Therefore, the proposed scheme can leverage the detection results among users in a severe wireless fading networks.
- The proposed spectrum sensing scheme uses a consensus algorithm to cope with two underlying network models, one with *fixed* bidirectional graphs and one with *random* graphs.

Our consensus-based approach is different from those used in distributed/decentralized detection problems [8–10,50]. In a typical distributed detection problem [8,9,50], each sensor individually forms its own discrete messages based on its local measurement and then reports to a fusion center, and there is in general no direct communication among the sensors. In certain models [10], a sensor may indirectly obtain information about other sensors, but this is achieved by feedback from a common fusion center.

## 3 Secondary Users Network Modeling

This section is organized in the following order. First, a network topology in distributed consensus-based cooperative spectrum sensing is presented. Then, the local spectrum sensing model is discussed in details. At last, the network model and consensus notions are presented.

### *3.1 Network Topology in Distributed Consensus-based Cooperative Spectrum Sensing*

As shown in Fig. 2, no common receiver is necessary compared with Fig. 1, and secondary users are communicating with each other via communication channels that are in good radio coverage of each of secondary users. Secondary users that are far away from each other do not have direct communication channels due to poor radio signal quality.

There are two stages in the proposed cognitive radio consensus schemes. In the first stage, secondary users use a spectrum sensing model to make measurements about primary users at the beginning of detection. This is done via sensing channels in Fig. 2. We denote the local measurement of user $i$ as $Y_i$. In the second stage, secondary users establish communication links with their own neighbors to locally exchange information among them, and then calculate the obtained data so as to make a local decision whether primary users are around. The above process in the second stage is done iteratively. At the initial time instant $k=0$, each user $i$ sets $x_i(0) = Y_i$ as the initial value of the local state variable. Next, at time $k = 0,1,2,\cdots$, according to the real-time network topology (or local wireless neighborhood), users mutually transmit and receive their states and then use local computation rules to generate updated states $x_i(k+1)$. Those iterations are done repeatedly until all the individual states $x_i(k)$ converge toward a common value $x^*$.

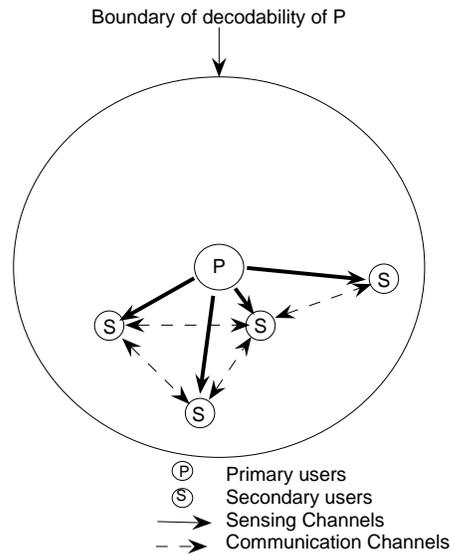

**Fig. 2:** A topology of distributed consensus-based cooperative spectrum sensing.

Before we introduce the detailed algorithms used in our consensus scheme, the common spectrum sensing model used in the first stage and the network model used in the second stage are to be presented, followed by the formal definition of the spectrum sensing consensus scheme.

## *3.2 The Spectrum Sensing Model*

In the first stage, secondary users make measurements about primary users at the beginning of each time slot. Three kinds of methods are widely used for spectrum sensing [6]: matched filter, energy detector and cyclostationary feature detector.

- **Matched Filter**

  The optimal way for any signal detection is a matched filter [51], since it maximizes received signal-to-noise ratio. However, a matched filter effectively requires demodulation of a primary user signal. This means that cognitive radio has a priori knowledge of primary user signal at both PHY and MAC layers [23, 25, 26, 30], e.g. modulation type and order, pulse shaping, packet format. Such information might be pre-stored in CR memory, but the cumbersome part is that for demodulation it has to achieve coherency with primary user signal by performing timing and carrier synchronization, even channel equalization. This is still possible since most primary users have pilots, preambles, synchronization words or spreading codes that can be used for coherent detection. For examples: TV signal has narrowband pilot for audio and video carriers; CDMA systems have dedicated spreading codes for pilot and synchronization channels; OFDM packets have preambles for packet acquisition. The main advantage of matched filter is that due to coherency it requires less time to achieve high processing gain [52]. However, a significant drawback of a matched filter is that a cognitive radio would need a dedicated receiver for every primary user class.

- **Energy Detector**

  One approach to simplify matched filtering approach is to perform non-coherent detection through energy detection. This sub-optimal technique has been extensively used in radiometry. There are several drawbacks of energy detectors that might diminish their simplicity in implementation. First, a threshold used for primary user detection is highly susceptible to unknown or changing noise levels. Even if the threshold would be set adaptively, presence of any in-band interference would confuse the energy detector. Furthermore, in frequency selective fading it is not clear how to set the threshold with respect to channel notches. Second, energy detector does not differentiate between modulated signals, noise and interference. Since, it cannot recognize the interference, it cannot benefit from adaptive signal processing for canceling the interferer. Furthermore, spectrum policy for using the band is constrained only to primary users, so a cognitive user should treat noise and other secondary users differently. Lastly, an energy detector does not work for spread spectrum signals: direct sequence and frequency hopping signals, for which more sophisticated signal processing algorithms need to be devised. In general, we could increase detector robustness by looking into a primary signal footprint such as modulation type, data rate, or other signal feature.

- **Cyclostationary Feature Detection**

  Modulated signals are in general coupled with sine wave carriers, pulse trains, repeating spreading, hoping sequences, or cyclic prefixes which result in built-in periodicity. Even though the data is a stationary random process, these modulated signals are characterized as cyclostationary, since their statistics, mean and autocorrelation, exhibit periodicity. This periodicity is typically introduced intentionally in the signal format so that a receiver can exploit it for: parameter estimation such as carrier phase, pulse timing, or direction of arrival. This can then be used for detection of a random signal with a particular modulation type in a background of noise and other modulated signals.

In summary, Matched filter is optimal theoretically, but it needs the prior knowledge of the primary system, which means higher complexity and cost to develop adaptive sensing circuits for different primary wireless systems. Energy detection is suboptimal, but it is simple to implement and does not have too much requirement on the position of primary users. Cyclostationary feature detection can detect the signals with very low SNR, but it still requires some prior knowledge of the primary user [4].

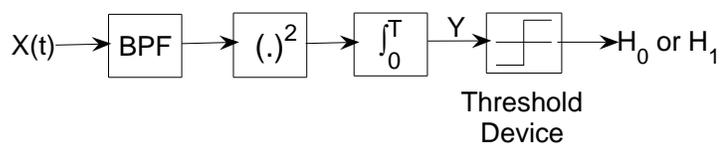

**Fig. 3:** Block diagram of an energy detector.

In this chapter, we consider the modeling scenario where the prior knowledge of the primary user is unknown. For implementation simplicity, an energy detection spectrum sensing method [5] is used. Fig. 3 shows the block-diagram of an energy detector. The input band pass filter (BPF) selects the center frequency $f_s$ and the bandwidth of interest $W$. This filter is followed by a squaring device and subsequently an integrator over a period of $T$. The output $Y$ of the integrator is the received energy at the secondary user and its distribution depends on whether the primary user signal is present or not. The goal of spectrum sensing is to decide between the following two hypotheses,

$$x(t) = \begin{cases} n(t), & H_0 \\ h \cdot s(t) + n(t), & H_1 \end{cases} \quad (2)$$

where $x(t)$ is the signal received by the secondary user, $s(t)$ is the primary user's transmitted signal, $n(t)$ is the additive white Gaussian noise (AWGN) and $h$ is the amplitude gain of the channel. We also denote by $\gamma$ the signal-to-noise ratio (SNR). The output of integrator in Fig. 3 is $Y$, which serves as the decision statistic. Following the work of [53], $Y$ has the following form,

$$Y = \begin{cases} \chi^2_{2TW}, & H_0 \\ \chi^2_{2TW}(2\gamma), & H_1 \end{cases} \quad (3)$$

where $\chi^2_{2TW}$ and $\chi^2_{2TW}(2\gamma)$ denote random quantities with central and non-central chi-square distributions, respectively, each with $2TW$ degrees of freedom and a non-centrality parameter of $2\gamma$ for the latter distribution. For simplicity we assume that the time-bandwidth product, $TW$, is an integer number, which is denoted by $m$.

Under Rayleigh fading, the gain $h$ is random, and the resulting SNR $\gamma$ would have an exponential distribution, so in this case the distribution of the output energy depends on the average SNR ($\bar{\gamma}$). When the primary user is absent, $Y$ is still distributed according to $\chi^2_{2TW}$. When the primary user is present, $Y$ may be denoted as the sum of two independent random variables [54], [55]:

$$Y = Y_\chi + Y_e, \qquad H_1, \tag{4}$$

where the distribution of $Y_\chi$ is $\chi^2_{2TW-2}$ and $Y_e$ has an exponential distribution with parameter $2(\bar{\gamma}+1)$.

As a summary, after $T$ seconds, each secondary user $i$ detects the energy and gets the measurement $Y_i \in \mathbb{R}^+$.

### 3.3 The Network Model and Consensus Notions

In the second stage, secondary users establish communication links with its neighbors to locally exchange information among them. In our scheme, the network formed by the secondary users can be described by a standard graph model. For simplicity, this can be represented by an undirected graph (to be simply called a graph) $\boldsymbol{G} = (\mathcal{N}, \mathcal{E})$ [56] consisting of a set of nodes $\{i = 1, 2, \cdots, n\}$ and a set of edges $\mathcal{E} \subset \mathcal{N} \times \mathcal{N}$. Denote each edge as an unordered pair $(i, j)$. Thus, if two secondary users are connected by an edge, it means they can mutually exchange information. A path in $\boldsymbol{G}$ consists of a sequence of nodes $i_1, i_2, \cdots, i_l$, $l \geq 2$, such that $(i_m, i_{m+1}) \in \mathcal{E}$ for all $1 \leq m \leq l-1$. The graph $\boldsymbol{G}$ is connected if any two distinct nodes in $\boldsymbol{G}$ are connected by a path. For convenience of exposition, we often refer node $i$ as secondary user $i$. The two names, secondary user and node, will be used interchangeably. The secondary user $j$ (resp., node $j$) is a neighbor of user $i$ (resp., node $i$) if $(j, i) \in \mathcal{E}$, where $j \neq i$. Denote the neighbors of node $i$ by $\mathcal{N}_i = \{j | (j, i) \in \mathcal{E}\} \subset \mathcal{N}$. The number of elements in $\mathcal{N}_i$ is denoted by $|\mathcal{N}_i|$ and called the degree of node $i$.

Throughout this chapter, the analysis is for undirected graphs, because we only deal with good duplex wireless links by which two adjacent nodes can establish communication (being connected) with each other. That is, the graph $\boldsymbol{G}$ is connected, and the information exchange between two neighboring nodes is bidirectional.

The Laplacian of the graph $\boldsymbol{G}$ is defined as $\boldsymbol{L} = (l_{ij})_{n \times n}$, where

$$l_{ij} = \begin{cases} |\mathcal{N}_i|, & \text{if } j = i \\ -1, & \text{if } j \in \mathcal{N}_i \\ 0, & \text{otherwise} \end{cases} \tag{5}$$

The matrix $\boldsymbol{L}$ defined by (5) is positive semi-definite. Further, if $\boldsymbol{G}$ is a connected undirected graph, then $\text{rank}(\boldsymbol{G}) = n - 1$ (see, e.g., [37]).

Since the cooperative spectrum sensing problem is viewed as a consensus problem where the users locally exchange information regarding their individual detection outcomes before reaching an agreement, we give the formal mathematical definition of consensus as follows.

The underlying network turns out to consist of secondary users reaching a consensus via local communication with their neighbors on a graph $\boldsymbol{G} = (\mathcal{N}, \mathcal{E})$.

For the $n$ secondary users distributed according to the graph model $\boldsymbol{G}$, we assign them a set of state variables $x_i$, $i \in \mathcal{N}$. Each $x_i$ will be called a consensus variable, and in the cooperative spectrum sensing context, it is essentially used by node $i$ for its estimate of the energy detection. By reaching consensus, we mean the individual states $x_i$ asymptotically converge to a common value $x^*$, i.e.,

$$x_i(k) \to x^* \text{ as } k \to \infty, \qquad (6)$$

for each $i \in \mathcal{N}$, where $k$ is the discrete time, $k = 0, 1, 2, \cdots$, and $x_i(k)$ is updated based on the previous states of node $i$ and its neighbors.

The special cases with $x^* = \text{Ave}(x) = (1/n)\sum_{i=1}^{n} x_i(0)$, $x^* = \max_{i=1}^{n} x_i(0)$ and $x^* = \min_{i=1}^{n} x_i(0)$ are called average-consensus, max-consensus, and min-consensus, respectively. It is worth mentioning that the existing spectrum sensing algorithm with the OR-rule can be viewed as a form of max-consensus. This chapter is intended to propose a cooperative spectrum sensing scheme in the framework of average-consensus.

## 4 Distributed Consensus-based Cooperative Spectrum Sensing in Fixed Graphs

In this chapter, let us assume the secondary users have established duplex wireless connections with their desired neighbors, and the connections remain working until the consensus is reached. This kind of topology is called as a fixed graph. Based on this assumption, we are going to propose the spectrum sensing consensus algorithm as follows.

### 4.1 The Consensus Algorithm

We denote for user $i$, its measurement $Y_i$ at time $k = 0$ by $x_i(0) = Y_i \in \mathbb{R}^+$. The state update of the consensus variable for each secondary user occurs at discrete time $k = 0, 1, 2, \cdots$, which is associated with a given sampling period. From $k = 0, 1, 2, \cdots$, the iterative form of the consensus algorithm can be stated as follows [37]:

$$x_i(k+1) = x_i(k) + \varepsilon \sum_{j \in \mathcal{N}_i} (x_j(k) - x_i(k)), \tag{7}$$

where

$$0 < \varepsilon < (\max_i |\mathcal{N}_i|)^{-1} \triangleq 1/\Delta. \tag{8}$$

The number $\Delta$ is called the maximum degree of the network.

This algorithm can be written in the vector form:

$$\boldsymbol{x}(k+1) = \boldsymbol{P}\boldsymbol{x}(k), \tag{9}$$

where $\boldsymbol{P} = \boldsymbol{I} - \varepsilon \boldsymbol{L}$. Notice that the upper bound in (8) for $\varepsilon$ ensures that $\boldsymbol{P}$ is a stochastic matrix, and in fact one can further show that $\boldsymbol{P}$ is ergodic when $\boldsymbol{G}$ is connected[1]. Since $\boldsymbol{G}$ is an undirected graph, all row sums and column sums of $\boldsymbol{L}$ are equal to zero. Hence $\boldsymbol{P}$ is a doubly stochastic matrix (i.e., $\boldsymbol{P}$ is a nonnegative matrix and all of its row sums and column sums are equal to one).

We also point out that (9) uses only a particular construction of the coefficient matrix for the consensus algorithm, which is based on the graph Laplacian $\boldsymbol{L}$. As long as each node has the prior knowledge of an upper bound of the maximum degree $\Delta$ of the network, the iteration may be implemented and there is no necessity for neighboring nodes to exchange information regarding the network structure. Also, it is possible to construct $\boldsymbol{P}$ in other forms. An alternative choice of $\boldsymbol{P}$ may be based on the so called Metropolis weights [46] by taking

$$\tilde{p}_{ij} = \begin{cases} \frac{1}{1+\max\{d_i, d_j\}} & \text{if } (j,i) \in \mathscr{E}, \\ 1 - \sum_{j \in \mathcal{N}_i} \tilde{p}_{ij} & \text{if } i = j, \\ 0 & \text{otherwise}, \end{cases}$$

where $d_i = |\mathcal{N}_i|$ is the degree of node $i$. If $\boldsymbol{G}$ is a connected graph and we define $\widetilde{\boldsymbol{P}} = (\tilde{p}_{ij})_{n \times n}$, then $\widetilde{\boldsymbol{P}}$ is an ergodic doubly stochastic matrix. When $\widetilde{\boldsymbol{P}}$ is used in (9) in place of $\boldsymbol{P}$, the state average will still be preserved as an invariant during the iterations and our convergence analysis below is still valid. Notice that when $\widetilde{\boldsymbol{P}}$ is used in the consensus algorithm, it is only required that any two neighboring nodes report to each other their degrees, and the knowledge of the maximum degree of the network is no longer needed.

---

[1] For some network topologies, it is possible to have an ergodic matrix $P = I - \varepsilon L$ when $\varepsilon = 1/\Delta$. For instance, if $\varepsilon$ is taken as $1/\Delta$ and meanwhile it is ensured that $P$ has at least one positive diagonal entry, then it can be shown that $P$ is an ergodic stochastic matrix.

We cite a theorem concerning the convergence property of the consensus algorithm.

**Theorem 1.** *(see, e.g., [37]) Consider a network of secondary users,*

$$x_i(k+1) = x_i(k) + u_i(k), \tag{10}$$

*with topology $\boldsymbol{G}$ applying the distributed consensus algorithm (7), where $u_i(k) = \varepsilon \sum_{j \in \mathcal{N}_i}(x_j(k) - x_i(k))$, $0 < \varepsilon < 1/\Delta$, and $\Delta$ is the maximum degree of the network. Let $\boldsymbol{G}$ be a connected undirected graph. Then*

1. *A consensus is asymptotically reached for all initial states;*
2. *$\boldsymbol{P}$ is doubly stochastic, and an average-consensus is asymptotically reached with the limit $x^* = (1/n)\sum_{i=1}^{n} x_i(0)$ for the individual states.* ∎

According to Theorem 1, if we choose $\varepsilon$ such that $0 < \varepsilon < 1/\Delta$, then an average-consensus is ensured and the final common value $x^* = (1/n)\sum_{i=1}^{n} x_i(0)$ will be the average of the initial vector $\boldsymbol{x(0)}$, or equivalently, the average of $\boldsymbol{Y}^T = \{Y_1, Y_2, \cdots, Y_n\}$, which has been obtained during the energy detection stage.

Finally, by comparing the average consensus result $x^*$ with a pre-defined threshold $\lambda$ based on Fig. 3, every secondary user $i$ gets the final data fusion locally:

$$\text{Decision } \boldsymbol{H} = \begin{cases} 1, & x^* > \lambda \\ 0, & \text{otherwise.} \end{cases} \tag{11}$$

## 4.2 Performance of the Consensus Algorithm

It is quite apparent that the convergence rate is yet another interesting issue in evaluating the performance of the spectrum sensing consensus algorithm. This is due to the fact that secondary users must continuously detect the presence of primary users, and back up as soon as possible on recognizing such incident. From this point of view, the speed of reaching a consensus is the key in the design of the network topology as well as the analysis of the performance of a consensus algorithm for a given spectrum sensing network. For the *connected* undirected graph $\boldsymbol{G}$, the above algorithm can ensure exponential convergence rate, where the error can be parameterized in the form $O(e^{-\delta t})$ with the exponent $\delta > 0$. To have some bound estimate for the parameter $\delta$, we first recall that $\boldsymbol{P} = \boldsymbol{I} - \varepsilon \boldsymbol{L}$. Since $L$ is a positive semi-definite matrix, denote its $n$ eigenvalues

by

$$0 = \lambda_1 < \lambda_2 \leq \cdots \leq \lambda_n. \tag{12}$$

Here $\lambda_2 > 0$ since the undirected graph **G** is *connected* which ensures that the rank of **L** is equal to $n-1$ ( [57]). The second smallest eigenvalue $\lambda_2$ of **L** is usually called the algebraic connectivity of the undirected graph **G**. Then the second largest absolute value of the eigenvalues of **P** is determined as $\alpha(\varepsilon) = \max\{|1 - \varepsilon\lambda_2|, |1 - \varepsilon\lambda_n|\}$, which can be verified to satisfy $\alpha(\varepsilon) < 1$. By using standard results in nonnegative matrix theory (see, e.g., [58]), we can obtain an upper bound for $\delta$. In fact, we can take $\delta$ as any value in the interval $(0, -\ln \alpha(\varepsilon))$. We also remark that similar convergence rate estimates can be carried out when general weight matrices in averaging are used.

Since **P** has a unit eigenvalue, we see that the difference between the first two largest absolute values of the eigenvalues of **P** is given as $g(\varepsilon) = 1 - \alpha(\varepsilon)$, which is customarily called the spectral gap of **P**. In general, the greater is $g(\varepsilon)$, the greater is the upper bound $-\ln \alpha(\varepsilon)$ for the exponent $\delta$, and the faster is the convergence of the consensus algorithm. In practical implementations, it is desirable to choose a suitable value for $\varepsilon$ to increase the spectral gap $g(\varepsilon)$ while **P** is ensured to be ergodic. We will discuss the convergence rate in the simulation part of this chapter.

## 5 Distributed Consensus-based Cooperative Spectrum Sensing in Random Graphs

In the previous section, it has been assumed that any two neighboring nodes can reliably exchange data at all times. Hence the network topology remains unchanged during the overall time period of interest. This kind of network modeling may not be accurate in certain situations. For example, fading of wireless signals can cause packet errors, which will result in wireless link failures for that period. Furthermore, even under LOS channels, moving objects between neighboring nodes may temporarily affect signal reception. For the above reasons, in this chapter, we consider a more realistic inter-node communication model with random link failures. Unlike the previous model, which is based on fixed bidirectional graphs, the new model is based on random graphs. Nevertheless, similar to the previous fixed topology scenario, for the random graph based modeling below, we still consider bidirectional links when two nodes can communicate.

## 5.1 Random Graph Modeling of the Network Topology

Before characterizing random connectivity of the network of all secondary users, let us first introduce a fixed undirected graph $\boldsymbol{G} = (\mathcal{N}, \mathcal{E})$ which describes the maximal set of communication links when there is no link failure. Due to the random link failures, at time $k$ the inter-user communication is described by a subgraph of $\boldsymbol{G}$ denoted by $\boldsymbol{G}(k) = (\mathcal{N}, \mathcal{E}(k))$ where $\mathcal{E}(k) \subset \mathcal{E}$; the edge $(j,i) \in \mathcal{E}(k)$ if and only if nodes $j$ and $i$ can communicate at time $k$ where $(j,i) \in \mathcal{E}$. Thus, the (undirected) graph $\boldsymbol{G}(k)$ is generated as the outcome of random link failures. Note that an edge $(j,i)$ never appears in $\boldsymbol{G}(k)$ if it is not an edge of $\boldsymbol{G}$. The neighbor set of node $i$ is $\mathcal{N}_i(k) = \{j | (j,i) \in \mathcal{E}(k)\}$ at time $k$. The number of elements in $\mathcal{N}_i(k)$ is denoted by $|\mathcal{N}_i(k)|$. At time $k \geq 0$, the adjacency matrix of $\boldsymbol{G}(k)$ is defined as $\boldsymbol{A}(k) = (\alpha_{ji}(k))_{1 \leq j, i \leq |\mathcal{N}|}$, where $\alpha_{ji}(k) = 1$ if $(j,i) \in \mathcal{E}(k)$, and $\alpha_{ji}(k) = 0$ otherwise. It is clear that the graph $\boldsymbol{G}(k)$ is completely characterized by the random matrix $\boldsymbol{A}(k)$.

Concerning the statistical properties of link failures, we assume that for all links (each associated with an edge in the graph $\boldsymbol{G}$) fail independently with the same probability $p \in (0,1)$. For notational simplicity we use the same parameter $p$ to model the failure probability. The generalization of the modeling and analysis to link-dependent failure probabilities is straightforward.

## 5.2 The Algorithm with Random Graphs

For the random link failure-prone model, the two spectrum sensing stages introduced in the previous chapter are still applicable. In the first stage, each node performs the radio detection and computes the measurements according to (2). During the second stage, at time $k$ each node exchanges states information with its neighbors and performs the corresponding computation to generate its state update $x_i(k+1)$. Let $\Delta$ be the maximum degree of the graph $\boldsymbol{G}$, and take $\varepsilon \in (0, 1/\Delta)$.

The state of user $i \in \mathcal{N}$ is updated by the rule

$$x_i(k+1) = x_i(k) + \varepsilon \sum_{j \in \mathcal{N}_i(k)} [x_j(k) - x_i(k)], \tag{13}$$

where $\varepsilon$ is a pre-determined constant step size. If $\mathcal{N}_i(k) = \emptyset$ (empty set), (13) reduces to $x_i(k+1) = x_i(k)$.

**Theorem 2.** *Under the independent link failure assumption, the algorithm (13) ensures average consensus, i.e., $\lim_{k \to \infty} x_i(k) = (1/n) \sum_{j=1}^{n} x_j(0)$ for all $i \in \mathcal{N}$, with probability one. If, in addition, $E|x(0)|^2 < \infty$ and* $x(0)$ is independent of the sequence of adjacency matrices $\mathbf{A}(k)$, $k = 0, 1, \cdots$, then each $x_i(k)$ converges to $(1/n)\sum_{j=1}^{n} x_j(0)$ in mean square with an exponential convergence rate.

Proof. We can write the algorithm (13) in the vector form

$$\boldsymbol{x}(k+1) = [\boldsymbol{I} - \varepsilon \boldsymbol{L}(k)]\boldsymbol{x}(k),$$

where $\boldsymbol{L}(k)$ is the Laplacian of the graph $\boldsymbol{G}(k)$. For a vector $\boldsymbol{z} = (z_1, \cdots, z_n)^T$, denote the Euclidean norm $|\boldsymbol{z}| = (\sum_{i=1}^{n} z_i^2)^{1/2}$. For any given sample point, we can show that $\boldsymbol{M}(k) = \boldsymbol{I} - \varepsilon \boldsymbol{L}(k)$ is a symmetric aperiodic stochastic matrix so that it has all its eigenvalues within the interval $(-1, 1]$ (see, e.g., [58]), and therefore $\boldsymbol{M}(k)$ determines a paracontracting map [46,59] in the sense $\boldsymbol{M}(k)\boldsymbol{z} \neq \boldsymbol{z}$ if and only if $|\boldsymbol{M}(k)\boldsymbol{z}| < |\boldsymbol{z}|$. For $\boldsymbol{M}(k)$, we denote its fixed point subspace $\mathscr{H}(\boldsymbol{M}(k)) = \{\boldsymbol{z} \in \mathbb{R}^n | \boldsymbol{M}(k)\boldsymbol{z} = \boldsymbol{z}\}$.

By the assumption on the independent link failures, we see that with probability one, $\boldsymbol{G}(k) = \boldsymbol{G}$ for an infinite number of times $k$. Let $\Omega$ denote the underlying probability sample space. Thus, after excluding a set $A_0$ of zero probability, for all $\omega \in \Omega \setminus A_0$, $\boldsymbol{G}(k) = \boldsymbol{G}$ infinitely often with the associated Laplacian being $\boldsymbol{L}(k) = \boldsymbol{L}$. Hence, for each $\omega \in \Omega \setminus A_0$, $x(k)$ converges to a point in the space $\mathscr{H}(\boldsymbol{I} - \varepsilon \boldsymbol{L}) = \{\boldsymbol{z} \in \mathbb{R}^n | \boldsymbol{L}\boldsymbol{z} = 0\}$ when $k \to \infty$. Furthermore, $\{\boldsymbol{z} \in \mathbb{R}^n | \boldsymbol{L}\boldsymbol{z} = 0\} = \text{span}\{\mathbf{1}_n\}$ since $\boldsymbol{G}$ is a connected undirected graph.

On the other hand, it is straightforward to check that $(1/n)\sum_{j=1}^{n} x_j(k)$ remains as a constant since $\boldsymbol{M}(k)$ is a doubly stochastic matrix (i.e., nonnegative matric with all row sums and column sums equal to one). Now it follows that each $x_i(k)$ converges to $(1/n)\sum_{j=1}^{n} x_j(0)$ with probability one, as $k \to \infty$.

We continue to analyze mean square convergence. Since $E|\boldsymbol{x}(0)|^2 < \infty$ and $\sup_{i \in \mathcal{N}, k \geq 0} |x_i(k)| \leq \max_{i \in \mathcal{N}} |x_i(0)| \leq |\boldsymbol{x}(0)|$, by the probability one convergence of $x_i(k)$, it follows from dominated convergence results in probability theory that $x_i(k)$ also converges to $(1/n)\sum_{j=1}^{n} x_j(0)$ in mean square.

Now, we proceed to give an estimation of the mean square convergence rate within the random network model. Denote $\text{Ave}(\boldsymbol{x}(0)) = (1/n)\sum_{j=1}^{n} x_j(0)$. It is straightforward to show that

$$\boldsymbol{x}(k+1) - \text{Ave}(\boldsymbol{x}(0))\mathbf{1}_n = [\boldsymbol{I} - (1/n)\mathbf{1}_n\mathbf{1}_n^T][\boldsymbol{I} - \varepsilon \boldsymbol{L}(k)][\boldsymbol{x}(k) - \text{Ave}(\boldsymbol{x}(0))\mathbf{1}_n] \quad (14)$$

$$\equiv \boldsymbol{B}(k)[\boldsymbol{x}(k) - \text{Ave}(\boldsymbol{x}(0))\mathbf{1}_n]. \quad (15)$$

In fact, for each $\omega \in \Omega$, by the eigenvalue distribution of the matrices $(1/n)\mathbf{1}_n\mathbf{1}_n^T$ and $\boldsymbol{L}(k)$, we can show that $\boldsymbol{B}^T(k)\boldsymbol{B}(k)$, and subsequently $E[\boldsymbol{B}^T(k)\boldsymbol{B}(k)]$, have $n$ real eigenvalues on the interval $[0, 1]$. We use a contradiction argument to show that the largest eigenvalue $\rho$ of $E[\boldsymbol{B}^T(k)\boldsymbol{B}(k)]$ is less than one. Suppose $\rho = 1$ for $E[\boldsymbol{B}^T(k)\boldsymbol{B}(k)]$; then there exists a real-valued vector $\boldsymbol{x} \neq 0$ such that

$$x^T E[B^T(k)B(k)]x = x^T x. \tag{16}$$

By the fact $x^T[B^T(k)B(k)]x \leq x^T x$, the equality (16) leads to

$$x^T[B^T(k)B(k)]x = x^T x \tag{17}$$

with probability one. On the other hand, by the link failure assumption, there exists a set $A_1 \subset \Omega$ such that $P(A_1) > 0$ and for each $\omega \in A_1$, the associated matrix value $B(k) = I - \varepsilon L$. Without the loss of generality, we can assume $A_1$ has been chosen in such a manner that for any $\omega \in A_1$, (17) also holds.

By noticing the fact that for any $z \in \mathbb{R}^n$,

$$z^T[B^T(k)B(k)]z \leq z^T(I - \varepsilon L)^2 z \leq z^T z, \tag{18}$$

we obtain from (17) that

$$x^T(I - \varepsilon L)^2 x = x^T x. \tag{19}$$

Hence, (19) implies that $x$ is the eigenvector of $I - \varepsilon L$ associated with the eigenvalue 1, which further implies that $x \in \text{span}\{1_n\}$. Denote $x = c1_n$ where $c$ is a constant. By substituting $x = c1_n$ into the left hand side of (17), we obtain $x^T[B^T(k)B(k)]x = 0$ for each $\omega \in \Omega$, which contradicts with (17) and the fact $x \neq 0$. Hence, we conclude that the largest eigenvalue $\rho$ of $E[B^T(k)B(k)]$ is in the interval $[0,1)$.

Finally, by elementary calculation we obtain the convergence rate estimate

$$E|x(k) - \text{Ave}(x(0))1_n|^2 \leq \rho^k E|x(0) - \text{Ave}(x(0))1_n|^2. \tag{20}$$

$\square$

In fact, we have the simplified expression:

$$\begin{aligned} B^T(k)B(k) &= [I - \varepsilon L(k)][I - (1/n)1_n 1_n^T]^2[I - \varepsilon L(k)] \\ &= [I - \varepsilon L(k)][I - (1/n)1_n 1_n^T][I - \varepsilon L(k)] \\ &= [I - \varepsilon L(k)]^2 - (1/n)1_n 1_n^T, \end{aligned}$$

and therefore, $\rho$ is also given as the largest eigenvalue of the positive semi-definite matrix $E[I - \varepsilon L(k)]^2 - (1/n)1_n 1_n^T$.

# 6 Simulation Results and Discussions

In this section, we present and discuss the simulation results of the distributed consensus-based scheme.

## *6.1 Distributed Consensus-Based Cooperative Spectrum Sensing*

### 6.1.1 Simulation Setup

In the simulations, we assume that all secondary users are experiencing i.i.d. Rayleigh fading without spatial correlation. Each secondary user uses an energy detector. We simulate the output $Y$ of the energy detector directly in our simulations. When the primary user is absent, $Y$ is a random quantity with chi-square distribution. When the primary user is present, $Y$ may be denoted as the sum of two independent random variables [54], [55]. The parameters of $Y$ depend on the average SNR in the Rayleigh fading (see (3) and (4)). The simulations are done in three test conditions. In the first condition, every user has the same average $SNR(\overline{\gamma})$, which is 10dB. In the second condition, each user has different average $SNR(\overline{\gamma})$ varying from 5dB to 9dB. In the third condition, each user has different average $SNR(\overline{\gamma})$ varying from 5dB to 15dB. The relevant information of primary users, such as the position, the moving direction and the moving velocity, is unknown to the secondary users.

We compare the performance of the proposed scheme with that of an existing OR-rule cooperative sensing scheme [29, 32, 60], which is better than AND-rule and MAJORITY-rule in many cases of practical interest [32, 60]. In the OR-rule cooperative sensing scheme, each secondary user makes local spectrum sensing decision, which is a binary variable - a "one" denotes the presence of a primary user, and a "zero" denotes its absence. Then, all of the local decisions are sent to a data collector to sum up all local decision values. If the sum is greater than or equal to one, a primary user is believed to be present.

In the first stage of spectrum sensing, after time synchronization, every secondary user performs energy detection with $TW = 5$ individually to get local measurement $Y_i$ at the selected center frequency $f_s$ and the bandwidth of interest $W$. To set up the initial energy vector $\mathbf{X}(0)$, we set $x_i(0) = Y_i$.

In the second stage, the existing method and the proposed consensus algorithm (7) are conducted based on fixed graph models, while the proposed consensus algorithm (13) is run based on random graph models. For fixed graphs, the basic requirement is to set up duplex wireless channels. In the simulations, we consider a network topology with 10 secondary users that establish a graph, $\mathbf{G} = \{\mathcal{N}, \mathcal{E}\}$, as shown in Fig. 0.4(a). For random graphs, we use the same set of nodes as in Fig. 0.4(b), but replace solid lines with dotted ones, which have probabilities of link failure of 40% (refer to Fig. 0.4(b)). The links in those figures stand for bidirectional wireless links. With regard to link failure probabilities, they mean both directions will fail to work in case of link failure. We also consider a network topology with 50 nodes in the simulations, which is shown in Fig. 5. All of the 50 nodes are located randomly. The links in the 50-node network have probabilities of failure of 40%.

### 6.1.2 Convergence of the Consensus Algorithm

Figs. 0.6(a) and 0.6(b) show the estimated primary user energy in the network with a 10-node fixed graph. We can observe that, although the initially sensed energy varies greatly due to their different wireless channel conditions for different secondary nodes, a consensus will be reached after several iterations. The step size $\varepsilon$ has effects on the convergence rate of the consensus algorithm. According to (7) and (13), a value should be selected for $\varepsilon$ such that $0 < \varepsilon < \Delta^{-1}$. Since the maximum number of neighbors of a node in Figs. 0.4(a) and 0.4(b) is 5, $\Delta = 5$. Then, $0 < \varepsilon < 0.2$.

Here we provide some discussion about the choice of the parameter $\varepsilon$. First, given the network topology, we may construct the associated Laplacian $\bm{L}$ as a $10 \times 10$ matrix. For reasons of space, $\bm{L}$ is not displayed. The eigenvalue of $\bm{L}$ are listed as follows:

$$0,\ 0.3416,\ 0.8400,\ 1.4239,\ 2.0000,\ 2.0000,\ 3.0000,\ 3.1373,\ 4.9411,\ 6.3161. \tag{21}$$

On the interval $(0,\ 0.2)$, the spectral gap $g(\varepsilon)$ may be shown to be

$$g(\varepsilon) = 1 - 0.3416\varepsilon, \tag{22}$$

which monotonically decreases on $(0,\ 0.2)$. We note that for this specific network topology, when $\varepsilon = 0.2$, the resulting matrix $\bm{P} = \bm{I} - \varepsilon \bm{L}$ is ergodic. On the interval $(0, 0.2]$ the spectral gap is maximized at $\varepsilon = 0.2$.

In below we select two values for $\varepsilon$, 0.1 and 0.19, in Fig. 0.6(a) and Fig. 0.6(b), respectively. We can see that the algorithm converges faster when $\varepsilon = 0.19$ than that when $\varepsilon = 0.1$, which is due to the fact that $\varepsilon = 0.19$ corresponds to a larger spectral gap $g(0.19)$.

After about 5 iterations in Fig. 0.6(b), the difference between the nodes is less than 1 dB, which indicates that a consensus is achieved. Fig. 7 shows the estimated primary user energy in the network with a random graph when $\varepsilon = 0.19$. Comparing Fig. 7 with Fig. 0.6(b), we can see that the algorithm converges more slowly in the random graph case due to the random link failure in the CR network. In Fig. 7, after about 10 iterations, the difference between the nodes is less than 1 dB, which indicates that a consensus is achieved.

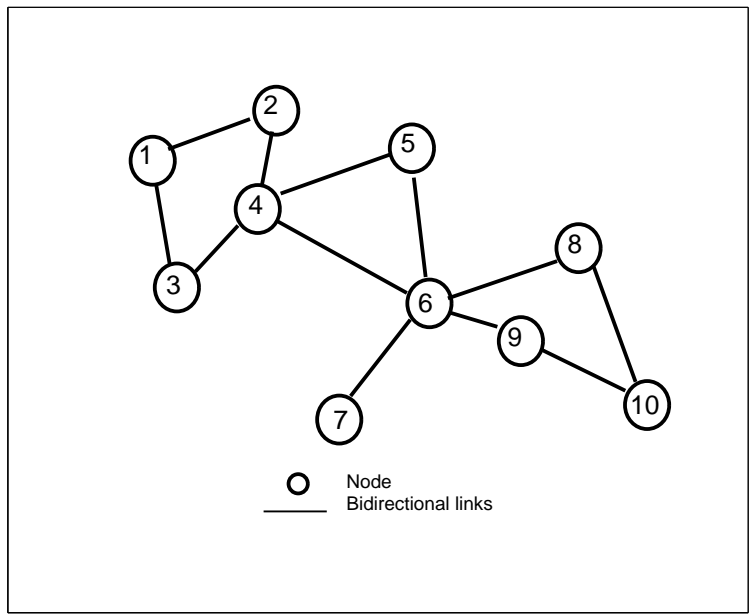

(a) A fixed graph.

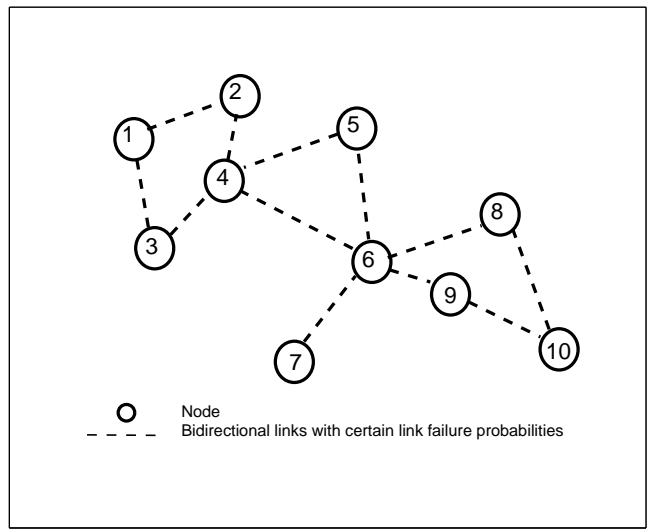

(b) A random graph.

**Fig. 4:** Network topology with 10 nodes in the simulations.

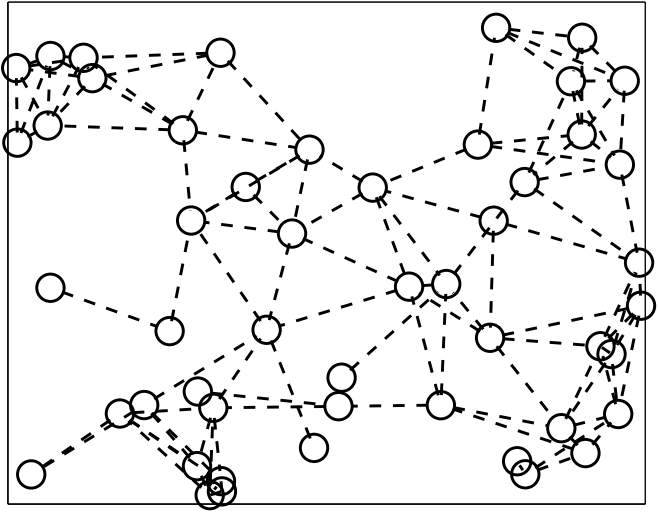

**Fig. 5:** Network topology with 50 nodes in the simulations.

Fig. 8 shows the convergence performance for the 50-node network. $\varepsilon = 0.15$ is used. We can observe that the algorithm converges more slowly in the 50-node network compared to the 10-node network due to a larger number of nodes. Nevertheless, after about 30 iterations, the difference between the nodes is less than 1 dB, which indicates that a consensus is achieved.

In the rest of the simulations, we conduct the simulations in three scenarios. In scenario one, under each of the three test conditions, the simulations are conducted by using one of the existing methods and the proposed scheme, respectively. The purpose of this scenario is to evaluate the performance of the proposed scheme in terms of $P_m$ (probability of missing detection) and $P_f$ (probability of false alarm). In scenario two, we focus on test condition one, and try to find the best detection sensitivity for different algorithms. In scenario three, we also work on test condition one, and set a fixed detection threshold $\lambda$ as stated in (11) to simulate the real situation in practice.

### 6.1.3 Scenario One

We compare the performance of the proposed scheme with that of an existing OR-rule cooperative sensing scheme [29, 32, 60]. Before the comparison, let us discuss briefly the relationship between $P_m$ (probability of missing detection)= 1 - $P_d$ (probability of detection) and $P_f$ (probability of false alarm). The fundamental tradeoff between $P_m$ and $P_f$ has different implications in the context of spectrum sensing [5]. A high $P_m$ will

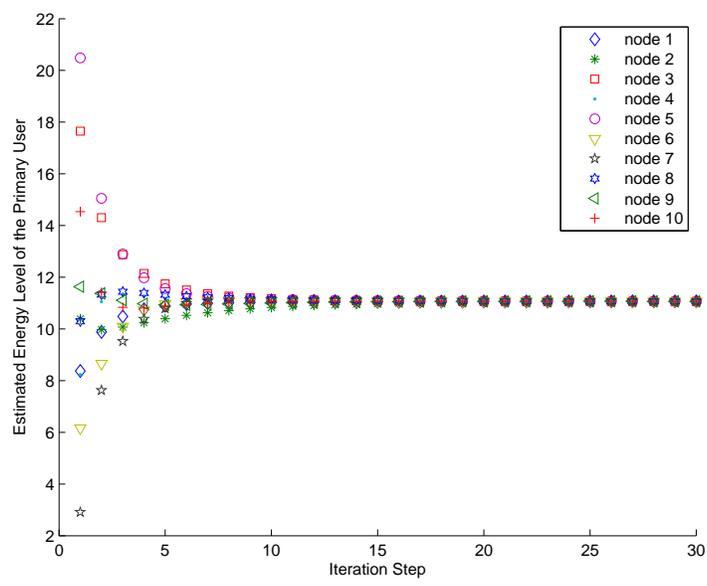

(a) Fixed graph ($\varepsilon = 0.1$).

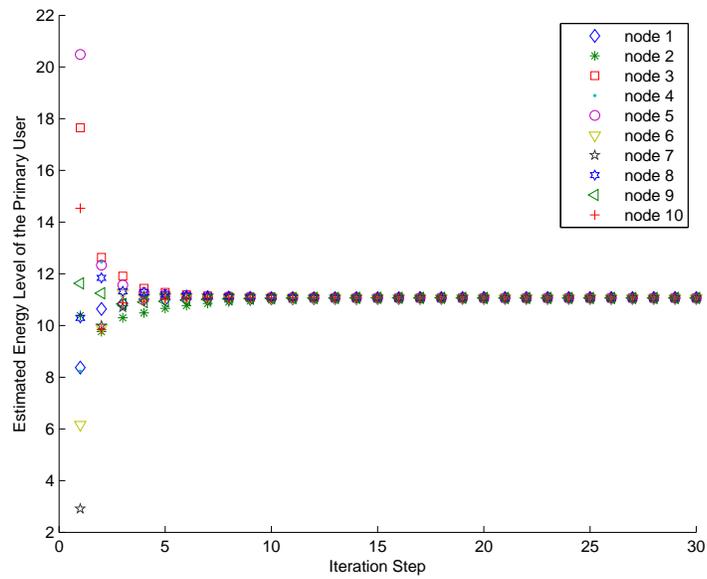

(b) Fixed graph ($\varepsilon = 0.19$).

**Fig. 6:** Convergence of the network with a 10-node fixed graph.

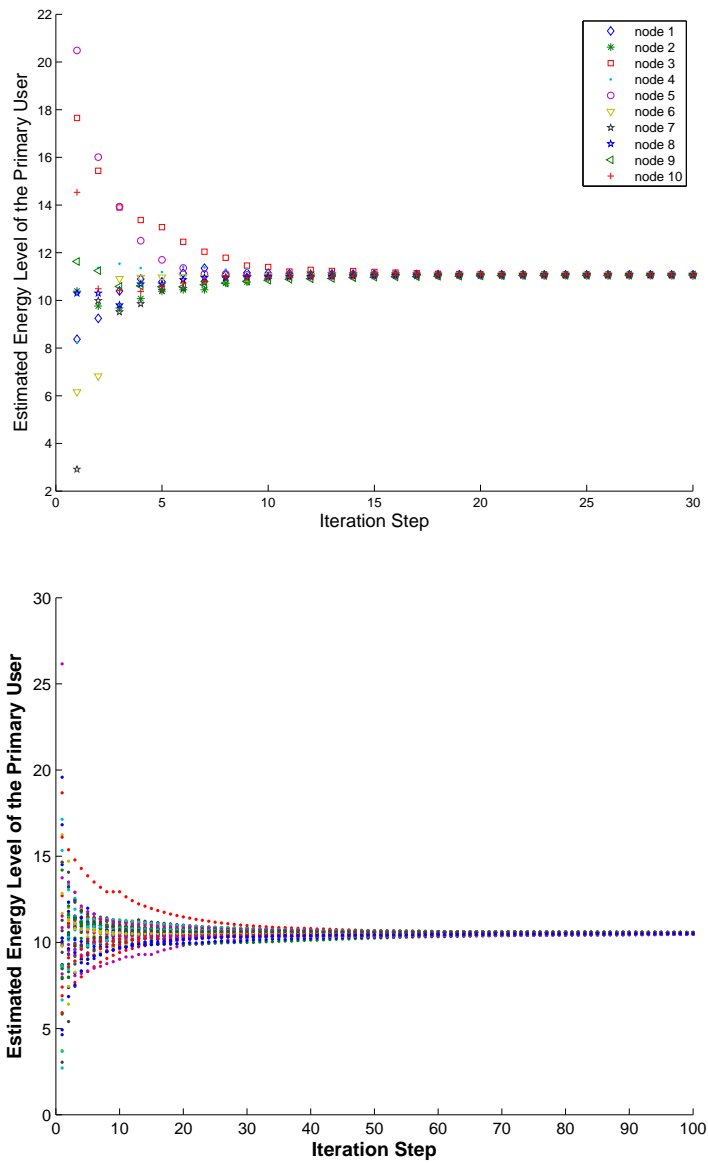

**Fig. 8:** Convergence of the network with a 50-node random graph ($\varepsilon = 0.15$).

result in the missing detection of primary users with high probability, which in turn increases the interference to primary users. On the other hand, a high $P_f$ will result in low spectrum utilization since false alarms increase the number of missed opportunities (white spaces). As expected, $P_f$ is independent of $\gamma$ since under $H_0$ there is no primary signal.

Figs. 9 and 10 show $P_f$ vs. $P_m$. We can see that the proposed algorithm has better performance than the existing OR-rule cooperative sensing scheme. The numbers beside the curves are the corresponding thresholds $\lambda$ in dB. In Fig. 9, where each secondary user has the same average SNR 10dB, if the threshold $\lambda$ is in the range of 11.4 to 12dB, both $P_f$ and $P_m$ can simultaneously drop below the probability of $10^{-2}$ for the proposed consensus algorithm in both fixed and random graphs. Also, the results are the same between the fixed and random models. In comparison, to reach the same goal, the existing OR-rule method must set $\lambda$ to be around 14.8dB, which has far worse $P_m$ ($10^{-2}$ vs. $10^{-3}$) with regard to the same $P_f$ level ($10^{-2}$).

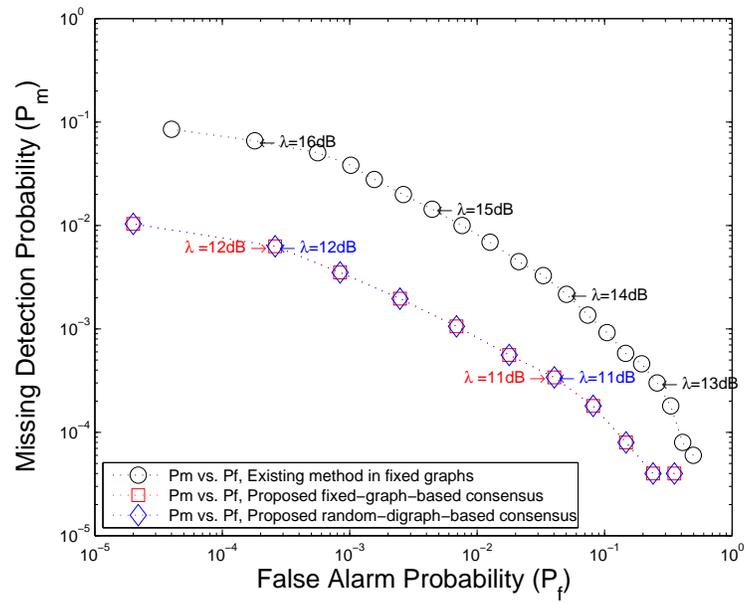

**Fig. 9:** Results in simulation scenario one under test condition one: Missing detection probability ($P_m$) vs. false alarm probability ($P_f$) (Each secondary user has the same average SNR, $\overline{\gamma} = 10$dB).

In condition two, secondary users undergo different average SNR varying from 5dB to 9dB. In condition three, secondary users undergo different average SNR varying from 5dB to 15dB. The similar results are demonstrated in Figs. 10 and 11 for condition two and three, respectively.

### 6.2 Scenario Two

Next, we examine the performance of detection probabilities $P_d$ to find out the sensitivity in detecting the primary user's presence. Fig. 12 shows $P_d$ (detection probability = $1 - P_m$) vs. average SNR ($\bar{\gamma}$) of secondary users. Condition one is used in this scenario, and the simulation is performed when the average SNR varies from 5dB to 10dB for all the nodes. The decision threshold, $\lambda$, is chosen so as to keep $P_f = 10^{-1}$. Time-bandwidth product, $TW$, is set to be 5, which is the same as before. From Fig. 12, we see that the proposed scheme can have a significant improvement in terms of the required average SNR for detection. In particular, if the probability of detection is expected to be kept above 0.99 (or $P_m < 10^{-2}$), the existing spectrum sensing scheme requires $\bar{\gamma} = 7.8$dB. This required average SNR is higher than those in the proposed consensus scheme, both of which are approximately 6.8dB.

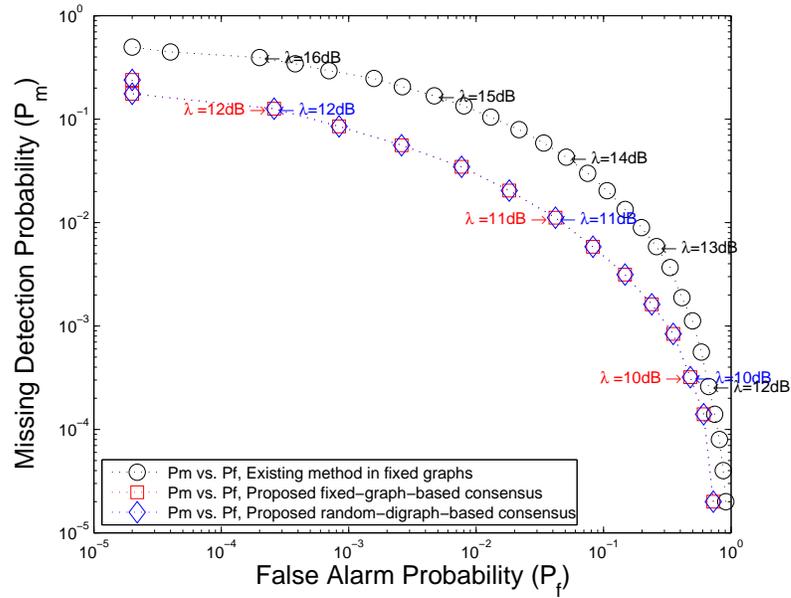

**Fig. 10:** Results in simulation scenario one under test condition two: Missing detection probability ($P_m$) vs. false alarm probability ($P_f$) (Each secondary user has different average SNR varying from 5dB to 9dB).

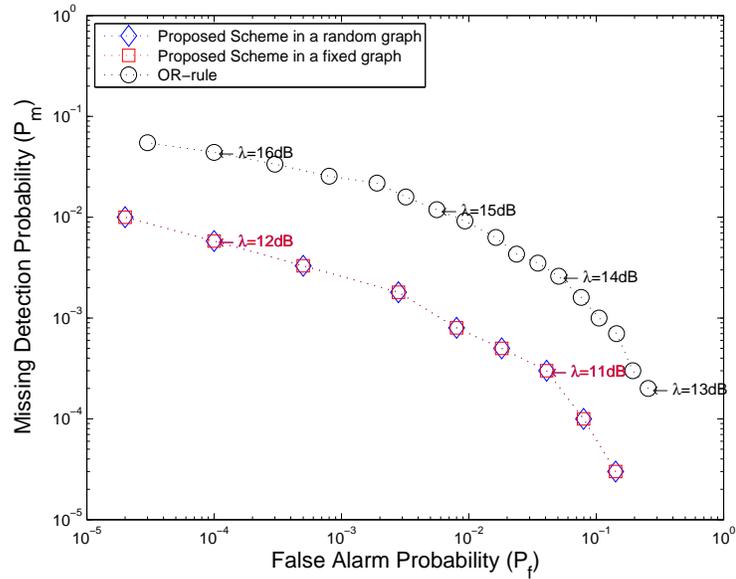

**Fig. 11:** Results in simulation scenario one under test condition three: Missing detection probability ($P_m$) vs. false alarm probability ($P_f$) (Each secondary user has different average SNR varying from 5dB to 15dB).

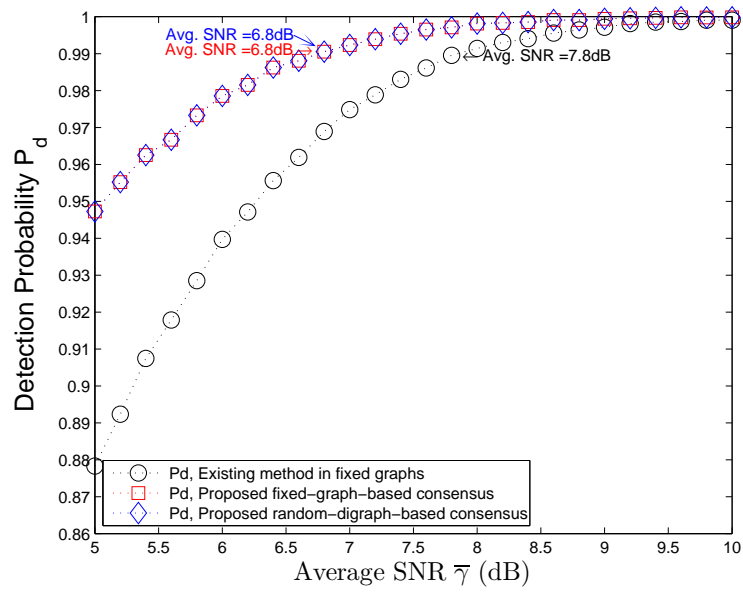

**Fig. 12:** Simulation results in scenario two: detection probability ($P_d$) vs. average SNR ($\bar{\gamma}$) ($P_f = 10^{-1}$, $TW = 5$).

### 6.2.1 Scenario Three

In reality, it is unlikely to adjust the threshold $\lambda$ on demand with regard to the different average SNR. Rather, a fixed threshold that can work in any $\bar{\gamma}$ is much more desirable. We can call it as threshold robustness. Therefore, in this scenario, we use condition one and intend to set a pre-defined threshold $\lambda$ by using (11) so as to achieve a certain goal. In fact, there are three options when we choose such a goal to keep missing detection probability ($P_m$) below a certain level, to keep false alarm probability $P_f$ around a certain level, or to keep both $P_m$ and $P_f$ as low as possible.

We first try to keep $P_m$ below $10^{-2}$ when all the ten users undergo the same $\bar{\gamma}$ varying from 5dB to 10dB. Fig. 0.13(a) shows a fixed $\lambda$ that lets $P_m$ below $10^{-2}$ for the average SNR ranging from 5dB to 10dB. As the result, the worst $P_f$ decreases from 0.586 by using the existing method to 0.356 in both the random graph and the fixed graph by using the proposed scheme.

The second option is to let $P_f$ always around $10^{-1}$ when all the ten users undergo $\bar{\gamma}$ varying from 5dB to 10dB. The result is shown in Fig. 0.13(b), where $P_f$ keeps around $10^{-1}$. The proposed consensus algorithm

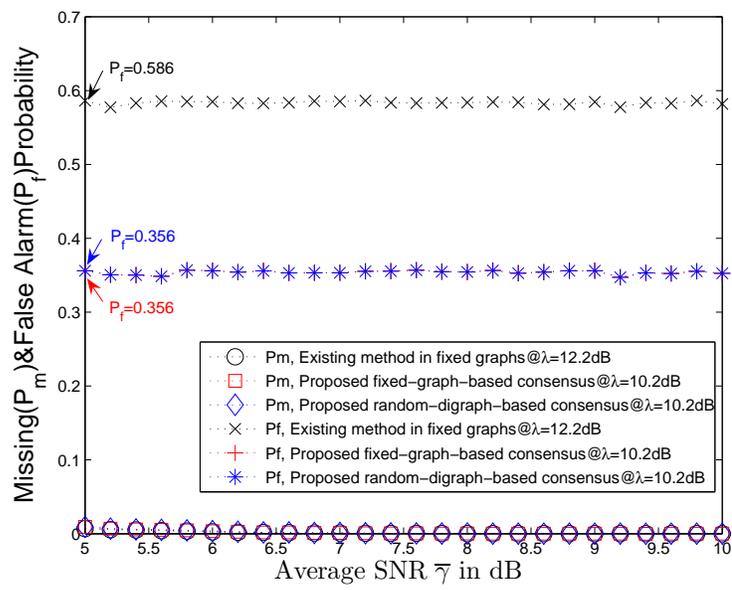

(a) Missing detection probability ($P_m$) and false alarm probability ($P_f$) vs. average SNR ($\bar{\gamma}$) with fixed threshold $\lambda$ to keep $P_m$ below $10^{-2}$, when all the ten users undergo same $\bar{\gamma}$ varying from 5dB to 10dB.

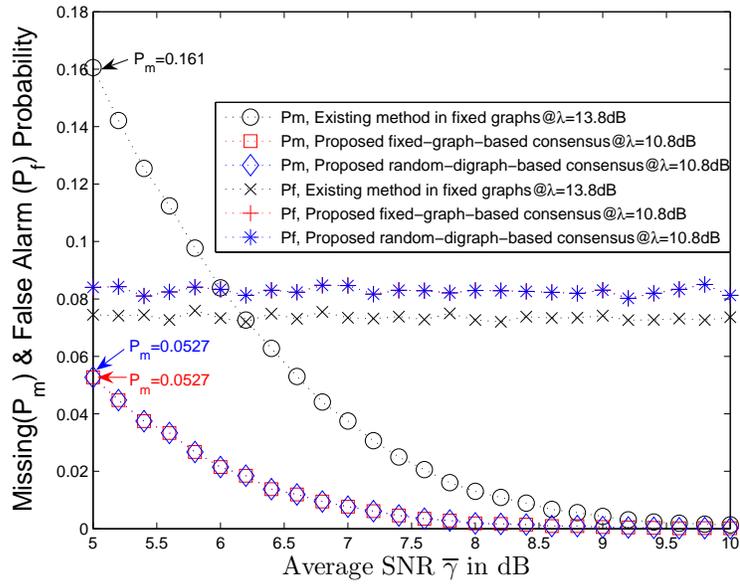

(b) Missing detection probability ($P_m$) and false alarm probability ($P_f$) vs. average SNR ($\bar{\gamma}$) with fixed threshold $\lambda$ to keep $P_f$ below $10^{-1}$, when all the ten users undergo same $\bar{\gamma}$ varying from 5dB to 10dB.

**Fig. 13:** Results in simulation scenario three: Part One.

has the better performance in terms of $P_m$, down from 0.161 in the existing method to 0.0527 in the proposed method.

In the third option, keep both $P_m$ and $P_f$ as low as possible. When determining a threshold, we refer to Fig. 0.14(a), which shows the worst case when all the ten users suffers $\overline{\gamma} = 5$dB. For the consensus scheme to have better missing detection performance, the threshold chosen in the proposed scheme should be lower than that in the OR-rule scheme. In Fig. 0.14(a), we can see that, with the same missing detection probability, the threshold is lower in the proposed scheme than that in the OR-rule scheme. On the other hand, with this lower threshold, a better false alarm probability can be achieved in the proposed scheme. The reason is that, when there is no primary user, the output of the energy detector, $Y$, of each secondary user is a random quantity with central chi-square distribution (please see Eq. (2)). Since $Y$ varies greatly, it is easy for a secondary user to have a false alarm in the OR-rule scheme. By contrast, the consensus scheme does not use the raw data $Y$ to make decisions. Instead, it uses the consensus among the secondary users to make decisions, thus it can remove some randomness in the raw data $Y$. Therefore, the consensus scheme can have a better false alarm probability than the OR-rule scheme with the same threshold. This can be shown in Fig. 0.14(a). From Fig. 0.14(a), we can also observe that both missing detection and false alarm probabilities are low when the threshold is round 11dB for the consensus scheme and when the threshold is around 13.6 dB for the OR-rule scheme. In Fig. 0.14(a), if we compare the performance of the consensus scheme with a threshold 11dB to that of the OR-rule scheme with a threshold 13.6 dB, we can see that both missing detection and false alarm probabilities are lower in the consensus scheme than those in the OR-rule scheme. We choose $\lambda = 11$dB for the proposed consensus algorithm, and $\lambda = 13.6$dB for the existing method to conduct our numerical studies. Fig. 0.14(b) illustrates the result of such a fixed $\lambda$. It is seen that both $P_m$ and $P_f$ have better performance for the proposed algorithm than those of the existing method. $P_m$ and $P_f$ drops to a relatively low level. This highlights the overall advantage in so-called threshold robustness for the proposed consensus algorithm. That is, for a given $\lambda$, the proposed consensus algorithm can output less $P_m$ and $P_f$ than those of the existing method. The algorithm works well in both fixed graphs and random ones. Another observation in scenario three is, when the average SNR rises, $P_m$ drops for a given threshold $\lambda$, but $P_f$ remains more or less at the same level. This means, for a fixed $\lambda$, $P_m$ is subject to the change of the average SNR. In contrast, $P_f$ is stable, because this parameter deals with the condition of $H_0$, where only the collective noises exists.

## 7 Conclusion

In this chapter, we have presented a fully distributed and scalable scheme for spectrum sensing based on recent advances in consensus algorithms. Cooperative spectrum sensing is modeled as a multi-agent coordination problem. Secondary users can maintain coordination based on only local information exchange

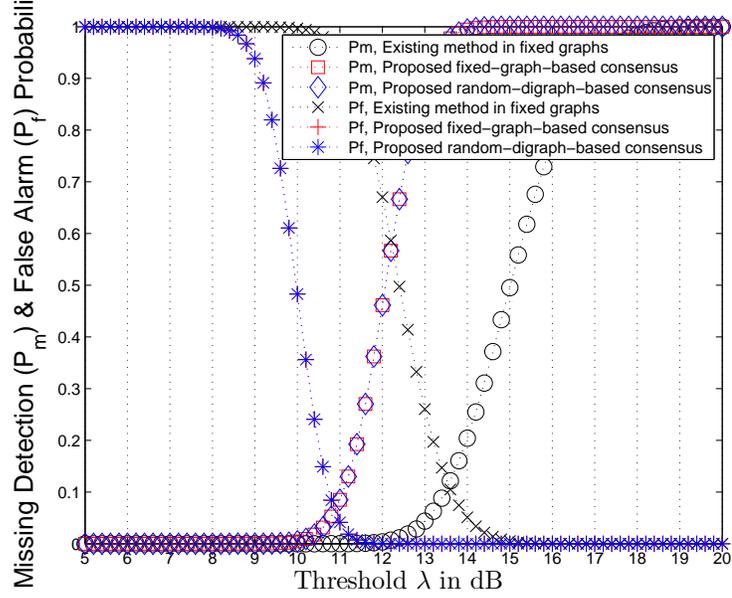

(a) Missing detection probability ($P_m$) and false alarm probability ($P_f$) vs. threshold $\lambda$ in dB when same $\overline{\gamma}$ = 5dB for all users.

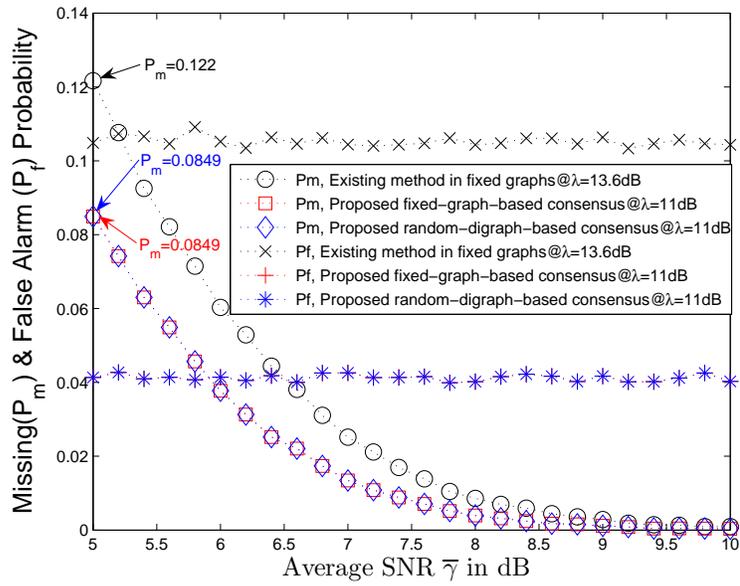

(b) Missing detection probability ($P_m$) and false alarm probability ($P_f$) vs. average SNR ($\overline{\gamma}$) with fixed threshold $\lambda$ to keep both $P_m$ and $P_f$ below a certain level, when all the ten users undergo same $\overline{\gamma}$ varying from 5dB to 10dB.

**Fig. 14:** Results in simulation scenario three: Part Two.

without a centralized receiver. Simulation results were presented to show the effectiveness of the proposed consensus-based scheme. It is shown that both missing detection probability and false alarm probability can be significantly reduced in the proposed scheme compared to those in the existing schemes.

Also, as the real network topologies undergo random changes and the primary user may randomly enter and leave the network, a protocol is necessary to quickly decide when the consensus is considered to be practical reached. If the secondary users cannot efficiently form a decision in finite steps, the energy measurements obtained at the beginning may become obsolete. To address this finite time detection issue, in implementations a certain toleration threshold may be used by the users. A secondary user may stop the iteration if it finds the difference between the states of each neighbor and itself has fallen below the threshold. The choice of threshold depends on empirical studies. Our simulation indicates that the threshold may be chosen to be around a fraction of 1 dB or close to 1 dB.

One limitation of the proposed scheme is that the choice of the step size $\varepsilon$ depends on the maximum number of neighbors of a node in the network. In other words, each node needs to have the prior knowledge of an upper bound of the maximum degree of the network. To solve this problem, an alternative approach may be used, which is based on so called Metropolis weights [46]. This approach does not need the knowledge of the maximum degree of the network. Future work is in progress in this direction. In addition, we plan to study transport layer issues [63] and heterogeneous networks issues [?] in the proposed framework. Moreover, we also want to simplify the data format of detection statistics from each secondary user to save the wireless bandwidth. Finally addition, as energy detection does not work well for spread spectrum signals, other approaches will be studied to deal with such networks.

# List of References

[1] J. Mitola, *Cognitive radio: An integrated agent architecture for software defined radio*. Doctor of Technology Thesis, Royal Inst. Technol. (KTH), Stockholm, Sweden, 2000.

[2] G. Ganesan and Y. Li, "Cooperative spectrum sensing in cognitive radio, part I: two user networks," *IEEE Trans. Wireless Commun.*, vol. 6, pp. 2204–2213, June 2007.

[3] S. Haykin, "Cognitive radio: Brain-empowered wireless communications," *IEEE J. Sel. Areas Commun.*, vol. 23, pp. 201–220, Feb. 2005.

[4] C. Sun, W. Zhang, and K. B. Letaief, "Cluster-based cooperative spectrum sensing in cognitive radio systems," in *Proc. IEEE ICC'07*, pp. 2511–2515, 2007.

[5] A. Ghasemi and E. Sousa, "Collaborative spectrum sensing for opportunistic access in fading environments," in *Proc. IEEE DySPAN'05*, pp. 131–136, 2005.

[6] D. Cabric, S. Mishra, and R. Brodersen, "Implementation issues in spectrum sensing for cognitive radios," in *Proc. Thirty-Eighth Asilomar Conference on Signals, Systems and Computers*, vol. 1, pp. 772–776, 2004.

[7] J. Hillenbrand, T. Weiss, and F. Jondral, "Calculation of detection and false alarm probabilities in spectrum pooling systems," *IEEE Commun. Letters*, vol. 9, no. 4, pp. 349–351, 2005.

[8] J.-F. Chamberland and V. V. Veeravalli, "Wireless sensors in distributed detection applications," *IEEE Signal Proc. Mag.*, vol. 24, pp. 16–25, May 2007.

[9] R. Niu and P. Varshney, "Performance analysis of distributed detection in a random sensor field," *IEEE Trans. Signal Proc.*, vol. 56, no. 1, pp. 339–349, 2008.

[10] V. Veeravalli, "Decentralized quickest change detection," *IEEE Trans. Inform. Theory*, vol. 47, no. 4, pp. 1657–1665, 2001.

[11] S. Mishra, A. Sahai, and R. Brodersen, "Cooperative sensing among cognitive radios," in *Proc. IEEE ICC'06*, pp. 1658–1663, 2006.

[12] W. Ren, R. Beard, and E. Atkins, "A survey of consensus problems in multi-agent coordination," in *Proc. American Control Conference'05*, pp. 1859–1864, 2005.

[13] J. Mitola and G. Q. Maguire, "Cognitive radio: Making software radios more personal," *IEEE Pers. Commun.*, vol. 6, pp. 13–18, Aug. 1999.

[14] I. Akyildiz, W. Lee, M. Vuran, and S. Mohanty, "NeXt generation/dynamic spectrum access/cognitive radio wireless networks: a survey," *Computer Networks*, vol. 50, no. 13, pp. 2127–2159, 2006.

[15] G. Ganesan and Y. Li, "Cooperative spectrum sensing in cognitive radio - part II: multiuser networks," *IEEE Trans. Wireless Commun.*, vol. 6, pp. 2214–2222, June 2007.

[16] G. Ganesan and Y. G. Li, "Agility improvement through cooperative diversity in cognitive radio," in *Proc. IEEE GLOBECOM'05*, pp. 2505–2509, 2005.

[17] Z. Li, F. R. Yu, and M. Huang, "A Distributed Consensus-Based Cooperative Spectrum Sensing in Cognitive Radios," *IEEE Trans. Veh. Tech.*, vol. 59, no. 1, pp. 383-393, Jan. 2010.

36  F. Richard Yu, Helen Tang, Minyi Huang, Peter Mason, and Zhiqiang Li

[18] F. R. Yu, M. Huang, and H. Tang, "Biologically Inspired Consensus-based Spectrum Sensing in Mobile Ad Hoc Networks with Cognitive Radios," *IEEE Networks*, pp. 26-30, June 2010.

[19] E. Peh and Y.-C. Liang, "Optimization for cooperative sensing in cognitive radio networks," in *Proc. IEEE WCNC'07*, pp. 27–32, 2007.

[20] X. Zhang and K. G. Shin, "Markov-Chain Modeling for Multicast Signaling Delay Analysis," *IEEE/ACM Trans. Networking.*, vol. 12, no. 4, pp. 667–680, Aug. 2004.

[21] J. Tang and X. Zhang, "Cross-Layer Design of Dynamic Resource Allocation of Diverse QoS Guarantees for MIMO-OFDM Wireless Networks," in *Proc. IEEE International Symposium on a World of Wireless, Mobile, and Multimedia Networks (WoWMoM 2005)*, Taormina, Italy, June, 2005.

[22] X. Zhang and K. G. Shin, "Statistical Analysis of Feedback-Synchronization Signaling Delay for Multicast Flow Control," in *Proc. IEEE INFOCOM'2001*, Anchorage, Alaska, USA, Apr. 2001.

[23] H. Su and X. Zhang, "CREAM-MAC: An Efficient Cognitive Radio-EnAbled Multi-Channel MAC Protocol for Wireless Networks," in *Proc. IEEE International Symposium on a World of Wireless, Mobile, and Multimedia Networks (WoWMoM 2008)*, Newport Beach, California, USA, June 2008.

[24] J. Unnikrishnan and V. V. Veeravalli, "Cooperative sensing for primary detection in cognitive radio," *IEEE J. Sel. Topics Signal Proc.*, vol. 2, no. 1, pp. 18–27, 2008.

[25] J. Tang and X. Zhang, "Cross-Layer Resource Allocation Over Wireless Relay Networks for Quality of Service Provisioning," *IEEE J. Sel. Areas Commun.*, vol. 25, no. 4, pp. 645–657, May 2007.

[26] H. Su and X. Zhang, "Opportunistic MAC Protocols for Cognitive Radio Based Wireless Networks," in *Proc. IEEE the 41st Conference on Information Sciences and Systems (CISS 2007)*, John Hopkings University, Baltimore, MD, USA, Marc. 2007.

[27] Z. Quan, S. Cui, and A. H. Sayed, "Optimal linear cooperation for spectrum sensing in cognitive radio networks," *IEEE J. Sel. Topics Signal Proc.*, vol. 2, no. 1, pp. 28–40, 2008.

[28] Y.-C. Liang, Y. Zeng, E. Peh, and A. T. Hoang, "Sensing-throughput tradeoff for cognitive radio networks," *IEEE Trans. Wireless Commun.*, vol. 7, no. 4, pp. 1326–1337, 2008.

[29] R. Chen, J.-M. Park, and K. Bian, "Robust distributed spectrum sensing in cognitive radio networks," in *Proc. INFOCOM 2008. The 27th Conference on Computer Communications. IEEE*, pp. 1876–1884, 2008.

[30] J. Tang and X. Zhang, "Clustering-Based Multi-Channel MAC Protocols for QoS-Provisionings Over Vehicular Ad Hoc Networks," *IEEE Trans. Veh. Tech.*, vol. 56, no. 6, pp. 3309-3323, Nov. 2007.

[31] J. Tang and X. Zhang, "Transmit Selection Diversity with Maximal-Ratio Combining for Multicarrier DS-CDMA Wireless Networks Over Nakagami-m Fading Channels," *IEEE J. Sel. Areas Commun.*, vol. 24, no. 1, pp. 104–112, Jan. 2006.

[32] W. Zhang and K. Ben Letaief, "Cooperative communications for cognitive radio networks," *Proc. IEEE*, vol. 97, no. 5, pp. 878–893, 2009.

[33] C. S. R. Murthy and B. S. Manoj, *Ad Hoc Wireless Networks: Architectures and Protocols*. Upper Saddle River, NJ: Prentice Hall, 2004.

[34] T. Nakano and T. Suda, "Applying biological principles to designs of network services," *Appl. Soft Comput.*, vol. 7, no. 3, pp. 870–878, 2007.

[35] I. Carreras, I. Chlamtac, F. D. Pellegrini, and D. Miorandi, "Bionets: Bio-inspired networking for pervasive communication environments," *IEEE Trans. Veh. Tech.*, vol. 56, pp. 218–229, Jan. 2007.

[36] F. Dressler, Ö. B. Akan, and A. Ngom, "Guest Editorial - Special Issue on Biological and Biologically-inspired Communication," *Springer Trans. on Computational Systems Biology (TCSB)*, vol. LNBI 5410, Dec. 2008.

[37] R. Olfati-Saber, J. Fax, and R. Murray, "Consensus and cooperation in networked multi-agent systems," *Proc. IEEE*, vol. 95, no. 1, pp. 215–233, 2007.

[38] H. Su and X. Zhang, "Cross-Layer Based Opportunistic MAC Protocols for QoS Provisionings Over Cognitive Radio Wireless Networks," *IEEE J. Sel. Areas Commun.*, vol. 26, no. 1, pp. 118–129, Jan. 2008.

[39] J. Tang and X. Zhang, "Quality-of-Service Driven Power and Rate Adaptation Over Wireless Links," *IEEE Trans. Wireless Commun.*, vol. 6, no. 8, pp. 3058–3068, Aug. 2007.

[40] J.-M. Amé, J. Halloy, C. Rivault, C. Detrain, and J. L. Deneubourg, "Collegial decision making based on social amplification leads to optimal group formation," *Proc. Natl. Acad. Sci.*, vol. 103, no. 15, pp. 5835–5840, 2006.

[41] L. Conradt and T. J. Roper, "Consensus decision making in animals," *Trends in Ecology and Evolution*, vol. 20, pp. 449–456, Aug. 2005.

[42] T. Vicsek, "A question of scale," *Nature*, vol. 441, p. 421, May 2001.

[43] I. D. Couzin, "Collective cognition in animal groups," *Trends in Cognitive Sciences*, vol. 13, pp. 36–43, Dec. 2008.

[44] P. K. Visscher, "How self-organization evolves?," *Nature*, vol. 421, pp. 799–800, Feb. 2003.

[45] W. Ren and R. Beard, "Consensus seeking in multiagent systems under dynamically changing interaction topologies," *IEEE Trans. Auto. Control*, vol. 50, no. 5, pp. 655–661, 2005.

[46] L. Xiao, S. Boyd, and S. Lall, "A scheme for robust distributed sensor fusion based on average consensus," in *Proc. Fourth International Symposium on Information Processing in Sensor Networks*, pp. 63–70, 2005.

[47] M. Huang and J. H. Manton, "Stochastic consensus seeking with measurement noise: convergence and asymptotic normality," in *Proc. American Control Conference'08*, pp. 1337–1342, 2008.

[48] A. Shamir, "Identity based cryptosystems and signature schemes," in *Proc. CRYPTO'84*, Aug. 1984.

[49] Y. Zhao, W. Liu, W. Lou, and Y. Fang, "Securing mobile ad hoc networks with certificateless public keys," *IEEE Trans. Dependable and Secure Computing*, vol. 3, pp. 386–399, Oct.–Dec. 2006.

[50] W. Irving and J. Tsitsiklis, "Some properties of optimal thresholds in decentralized detection," *IEEE Trans. Auto. Control*, vol. 39, no. 4, pp. 835–838, 1994.

[51] J. Proakis and M. Salehi, *Digital communications*. McGraw-hill New York, 1995.

[52] A. Sahai, N. Hoven, and R. Tandra, "Some fundamental limits on cognitive radio," in *Allerton Conference on Communication, Control, and Computing*, Citeseer, 2004.

[53] H. Urkowitz, "Energy detection of unknown deterministic signals," *Proc. IEEE*, vol. 55, no. 4, pp. 523–531, 1967.

[54] F. Digham, M.-S. Alouini, and M. Simon, "On the energy detection of unknown signals over fading channels," in *Proc. IEEE ICC'03*, vol. 5, pp. 3575–3579, 2003.

[55] V. Kostylev, "Energy detection of a signal with random amplitude," in *IEEE Proc. ICC'02*, vol. 3, pp. 1606–1610, 2002.

[56] M. Huang and J. H. Manton, "Coordination and consensus of networked agents with noisy measurements: stochastic algorithms and asymptotic behavior," *SIAM J. Control and Optimization*, vol. 48, pp. 134–161, Jan. 2009.

[57] C. Godsil and G. Royle, *Algebraic Graph Theory*. New York: Springer-Verlag, 2001.

[58] E. Seneta, *Non-negative Matrices and Markov Chains*. New York: Springer-Verlag, 1981.

[59] L. Elsner, I. Koltracht, and M. Neumann, "On the convergence of asynchronous paracontractions with applications to tomographic reconstruction from incomplete data," *Linear Algebra and its Applications*, vol. 130, pp. 65–82, 1990.

[60] A. Ghasemi and E. Sousa, "Opportunistic spectrum access in fading channels through collaborative sensing," *Journal of Communications*, vol. 2, no. 2, p. 71, 2007.

[61] X. Zhang, K. G. Shin, D. Saha, and D. Kandlur, "Scalable Flow Control for Multicast ABR Services in ATM Networks," *IEEE/ACM Trans. Networking.*, vol. 10, no. 1, pp. 67–85, Feb. 2002.

[62] J. Tang and X. Zhang, "Cross-Layer Modeling for Quality of Service Guarantees Over Wireless Links," *IEEE Trans. Wireless Comm.*, vol. 6, no. 12, pp. 4504-4512, Dec. 2007.

[63] L. Ma, F. Yu, V. C. M. Leung, and T. Randhawa, "A New Method to Support UMTS/WLAN Vertical Handover Using SCTP," *IEEE Wireless Communications*, vol. 11, no. 4, pp. 44-51, Aug. 2004.

[64] F. Yu and V. Krishnamurthy, "Optimal Joint Session Admission Control in Integrated WLAN and CDMA Cellular Networks with Vertical Handoff," *IEEE Trans. Mobile Computing*, vol. 6, no. 1, pp. 126-139, Jan. 2007.

# List of References